\documentclass{webofc}

\usepackage[varg]{txfonts}   
\usepackage{hyperref}
\usepackage{url}
\usepackage{overpic}
\usepackage{xspace}

\newcommand{\NASixtyOne}{NA61\slash SHINE\xspace}
\newcommand{\byc}{\kern-0.1em/\kern-0.1em c}
\newcommand{\GeV}{\ensuremath{\mbox{Ge\kern-0.1em V}}\xspace}
\newcommand{\MeV}{\ensuremath{\mbox{Me\kern-0.1em V}}\xspace}
\newcommand{\GeVc}{\ensuremath{\mbox{Ge\kern-0.1em V}\byc}\xspace}
\newcommand{\AGeVc}{\ensuremath{A\,\mbox{Ge\kern-0.1em V}\byc}\xspace}
\newcommand{\MeVc}{\ensuremath{\mbox{Me\kern-0.1em V}\byc}\xspace}

\newcommand{\pp}{\mbox{\textit{p+p}}\xspace}

\newcommand{\snn}{\ensuremath{\sqrt{s_{\mathrm{NN}}}}\xspace}
\usepackage{comment}

\hypersetup{colorlinks=true,citecolor=black,urlcolor=black,linkcolor=black}
\begin{document}
\title{NA61/SHINE results on multiplicity and net-charge fluctuations at CERN SPS energies}
%
%

\author{\firstname{Maja} \lastname{Ma\'{c}kowiak-Paw{\l}owska}\inst{1}\fnsep\thanks{\email{maja.pawlowska@pw.edu.pl}} for the NA61/SHINE Collaboration  
}

\institute{Warsaw University of Technology, Faculty of Physics, Koszykowa 75, 00-662 Warsaw, Poland}

\abstract{
This contribution presents final results on the energy dependence of multiplicity and net-electric charge fluctuations in Ar+Sc interactions collected by the \NASixtyOne experiment. Fluctuations were analyzed using higher-order cumulant ratios in Ar+Sc most central collisions and compared to \pp interactions. The results consider detector bias and centrality selection in the case of Ar+Sc. The results are compared across different energies and against model predictions.
}

\maketitle
\section{Introduction}
\label{sec:intro}

The properties of strongly interacting matter cannot be measured directly as quarks are confined in hadrons; instead, they are inferred from hadron and ion collision observables.  
Due to the statistical nature of particle production, the phase diagram of strongly interacting matter is commonly represented in terms of temperature, $T$, and baryo-chemical potential, $\mu_B$~\cite{Stock:2009yg}. At lower values of $\mu_B$, one expects a smooth crossover transition from hadrons to deconfined quarks and gluons~\cite{HotQCD:2018pds}. At high $\mu_B$, a first-order phase transition is predicted from the models~\cite{Fischer:2014ata}.  Experimental results also suggest the presence of a phase transition in the lower collision energy range of the SPS accelerator~\cite{NA49:2007stj}. 
In such a case, the phase transition line should end with a critical point of the second order (CP)~\cite{Antoniou:2002xq}. The search for possible CP signals is one of the motivations of high-energy nuclear collision studies at different laboratories.
Among them, the \NASixtyOne within its strong interaction program aims to search for the CP signals and study the properties of the onset of deconfinement. For these reasons, it performs a system-size energy scan of different ion species (\pp, Be+Be, Ar+Sc, Xe+La, Pb+Pb, and $p$+Pb).
\NASixtyOne experiment is located in the North Area of CERN, utilizing beams from the SPS accelerator in the range of 13$A$--150$A$/158\AGeVc beam momentum (collision center-of-mass energy per nucleon pair \snn = 5.1--16.8/17.3~\GeV)~\cite{Antoniou:2006mh, Abgrall:2014fa}.

The search for possible CP signals is performed by studying fluctuations and correlations of hadrons produced in high-energy collisions. As the system approaches CP, multiplicity and transverse momentum distributions are expected to develop non-Gaussian features, such as double-peaked structures associated with phase coexistence~\cite{Stephanov2011QCDcritical}. The shape of the distribution can be quantified through its moments and cumulants. In general, the higher the order of the moment or cumulant ($\kappa$), the higher the sensitivity of it on correlation length $\xi$, e.g., $\kappa_{2}\sim\xi^2$ whereas $\kappa_{4}\sim\xi^{7}$. It should be stressed that the CP signal is not expected in \pp interactions due to the small system size of the created system.

\section{Results}
So far, the \NASixtyOne experiment investigated, among other observables relevant to the search for CP, the energy dependence of transverse momentum and multiplicity fluctuations in \pp, Be+Be, and Ar+Sc collisions \cite{Grebieszkow:2017gqx}; the femtoscopis analysis in Be+Be and Ar+Sc interactions at \snn = 5.1--16.8~\GeV~\cite{QM2025_KG, NA61SHINE:2023qzr};
the intermittency of protons in Ar+Sc at \snn = 5.1--16.8~\GeV \cite{NA61SHINE:2023gez, SHINE:2024xtq} and in Pb+Pb at \snn = 5.1 and 7.6~\GeV~\cite{Adhikary:2022sdh}; and the intermittency of negatively charged hadrons in Xe+La at \snn = 16.8~\GeV~\cite{ReynaOrtiz:2024hul, QM2025_KG} and in Pb+Pb at \snn = 7.6~\GeV~\cite{Adhikary:2023rfj}. 
These results do not indicate structures that could be interpreted as CP signatures. 

\begin{figure}[b]
    \centering
    \includegraphics[width=0.32\textwidth]{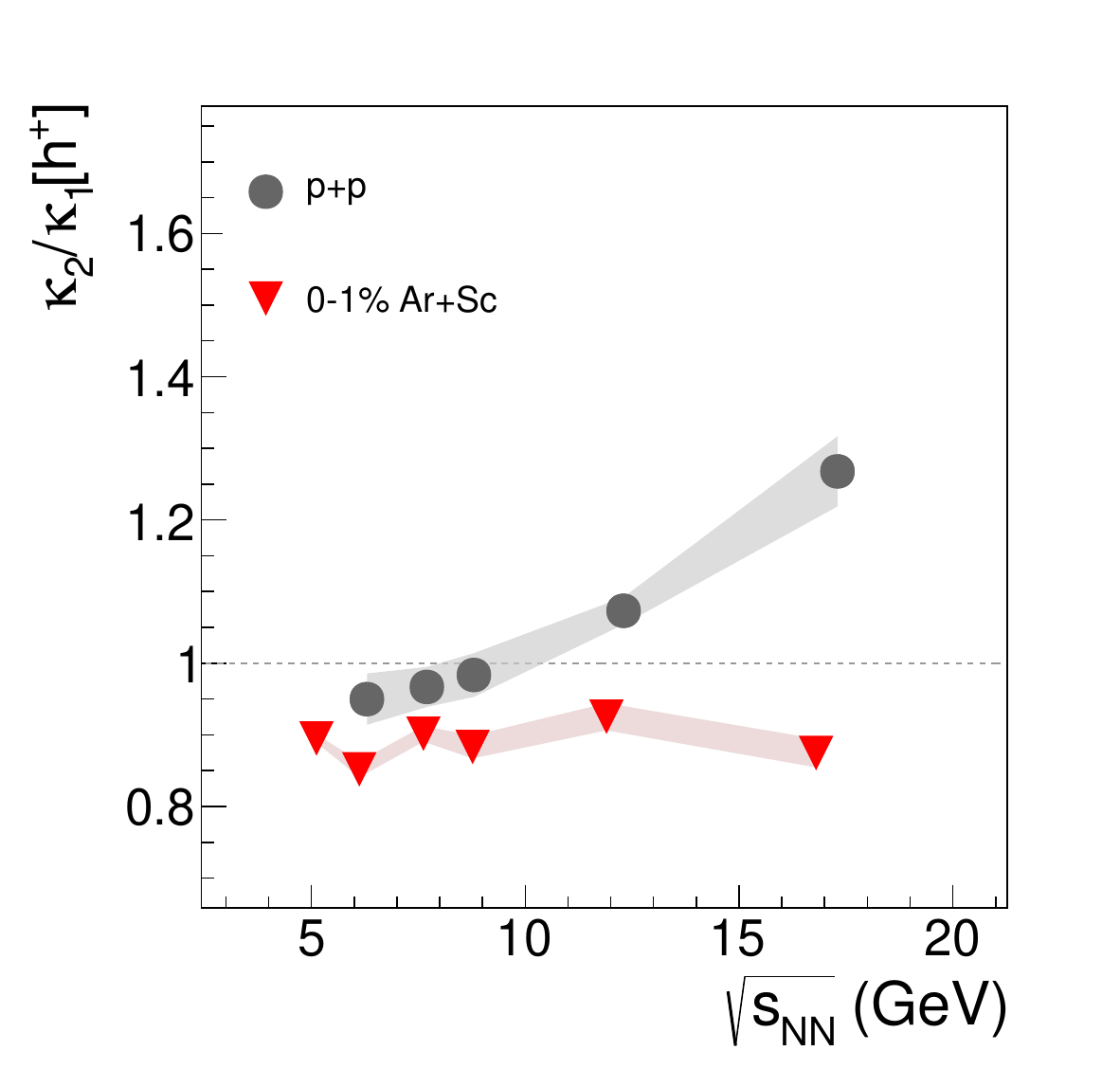}
    \includegraphics[width=0.32\textwidth]{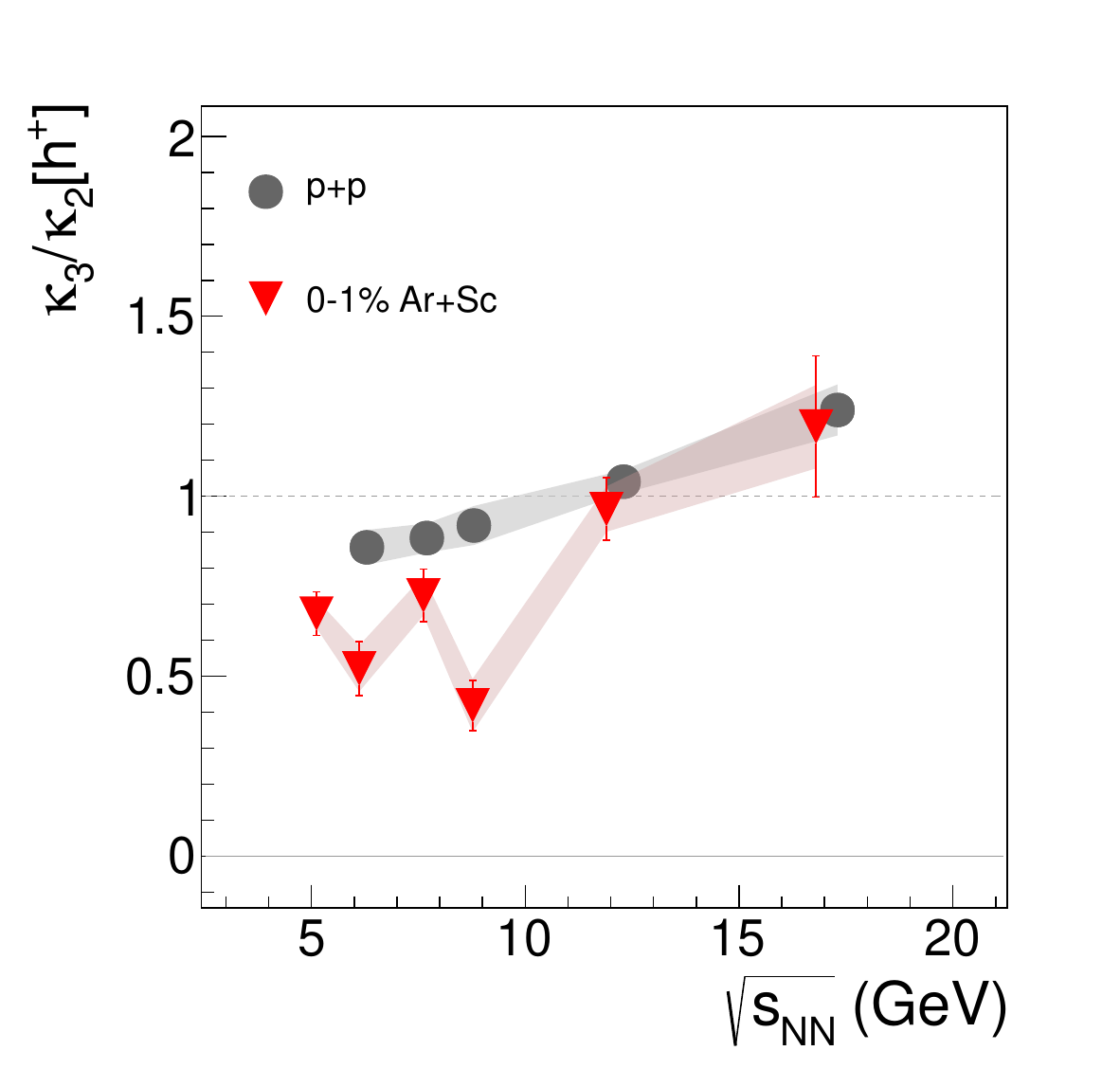}
    \includegraphics[width=0.32\textwidth]{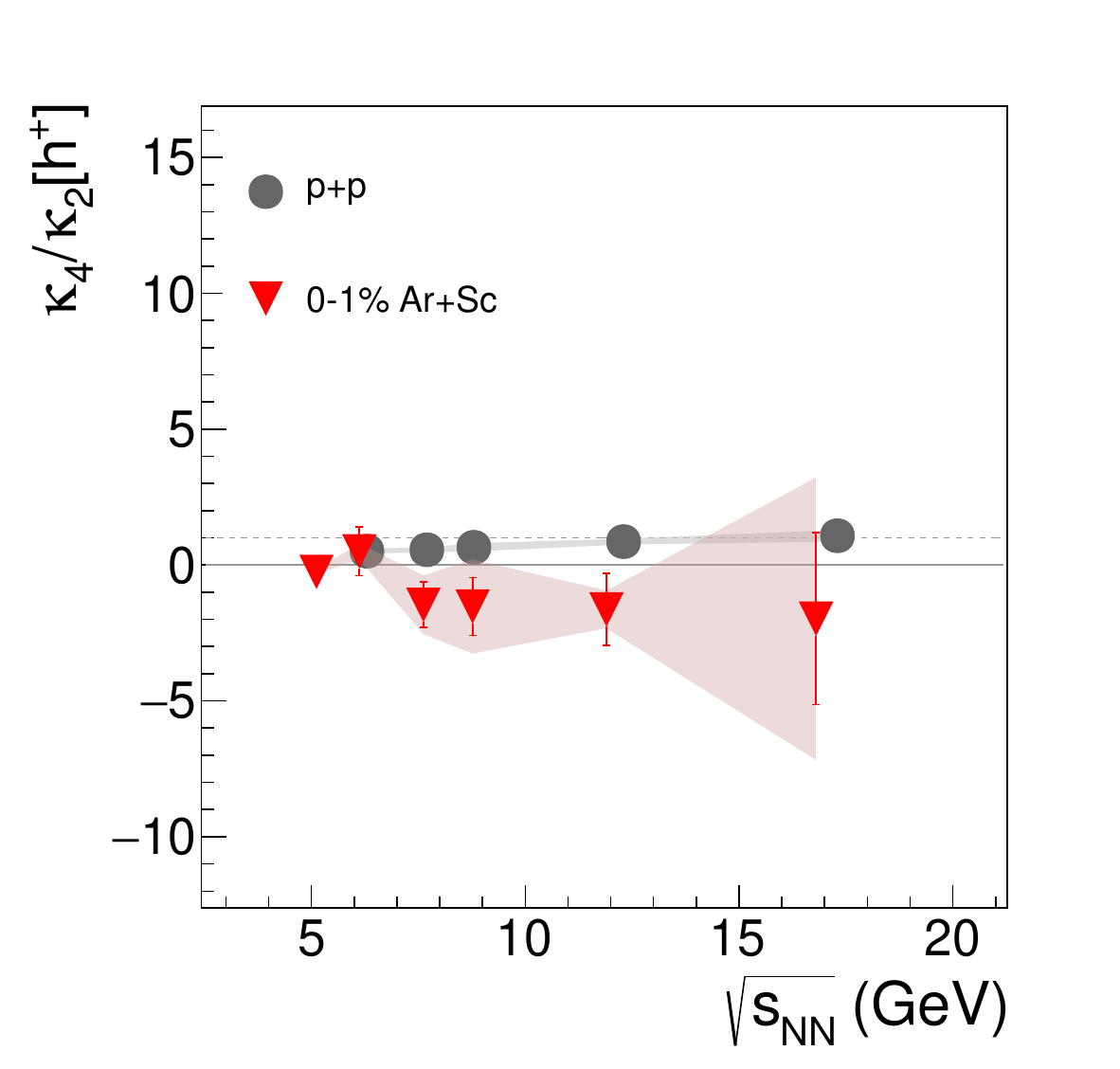}
    \caption{Energy dependence of intensive quantities of positively charged hadron cumulant ratios in Ar+Sc collisions compared to published \pp~\cite{SHINE:2023ejm}. 
Vertical bars denote statistical uncertainties, and color bands -- systematic ones. Gray lines indicate the reference values~\cite{NA61SHINE:2025whi}.}
    \label{fig:results_pos}
\end{figure}

This article focuses on new studies of multiplicity and net-electric charge (net-charge) fluctuations in the 1$\%$ most central $^{40}$Ar+$^{45}$Sc interactions at \snn = 5.1--16.8~\GeV collected by \NASixtyOne in 2015. Details on the measured energy dependencies and comparison with \pp interactions and EPOS1.99 model predictions can be found in Ref.~\cite{NA61SHINE:2025whi}.   
\begin{figure}
    \centering
    \begin{minipage}{0.37\textwidth}
    \includegraphics[width=\textwidth]{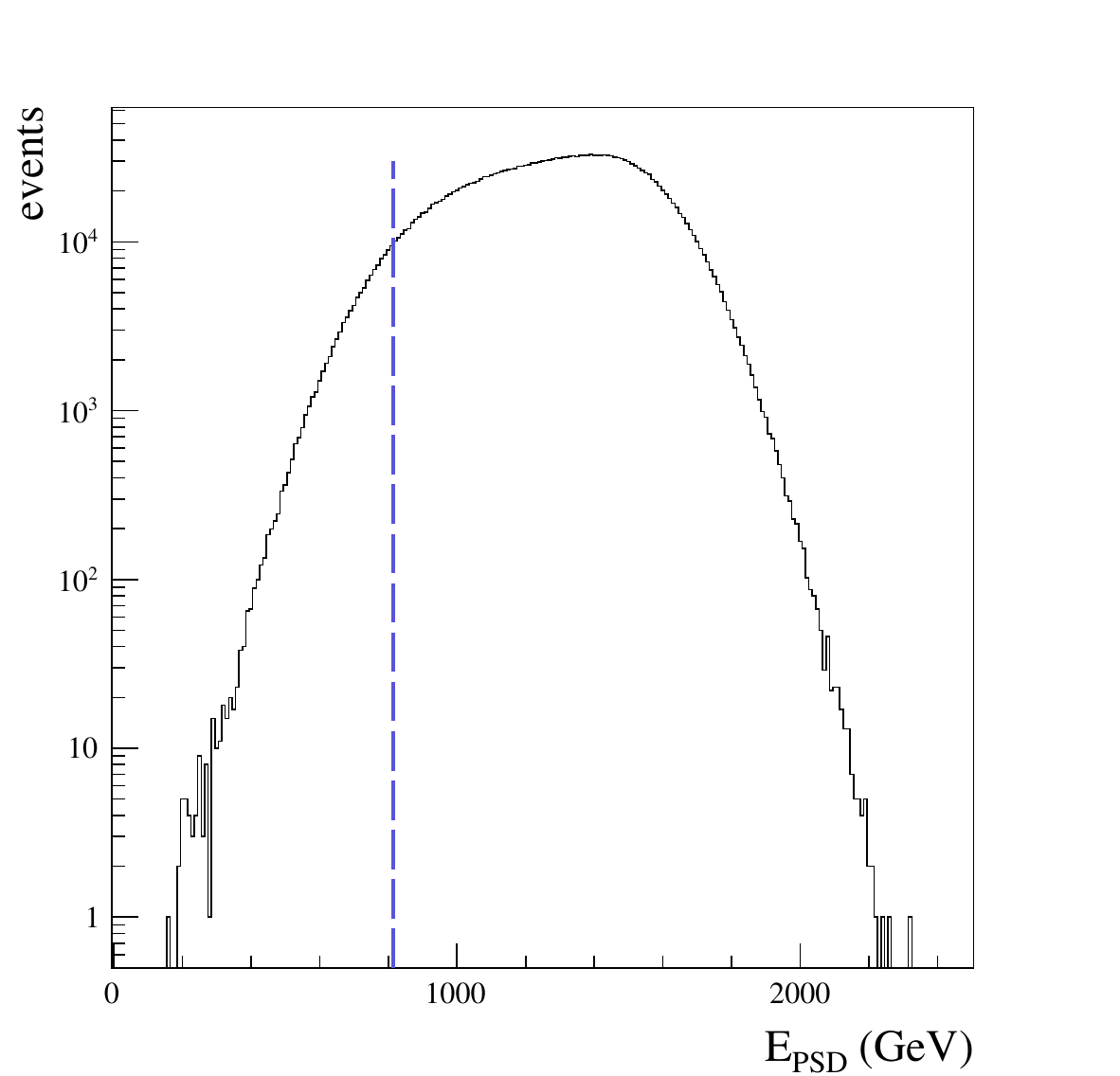}    
    \end{minipage}
    \begin{minipage}{0.4\textwidth}
    \includegraphics[width=\textwidth]{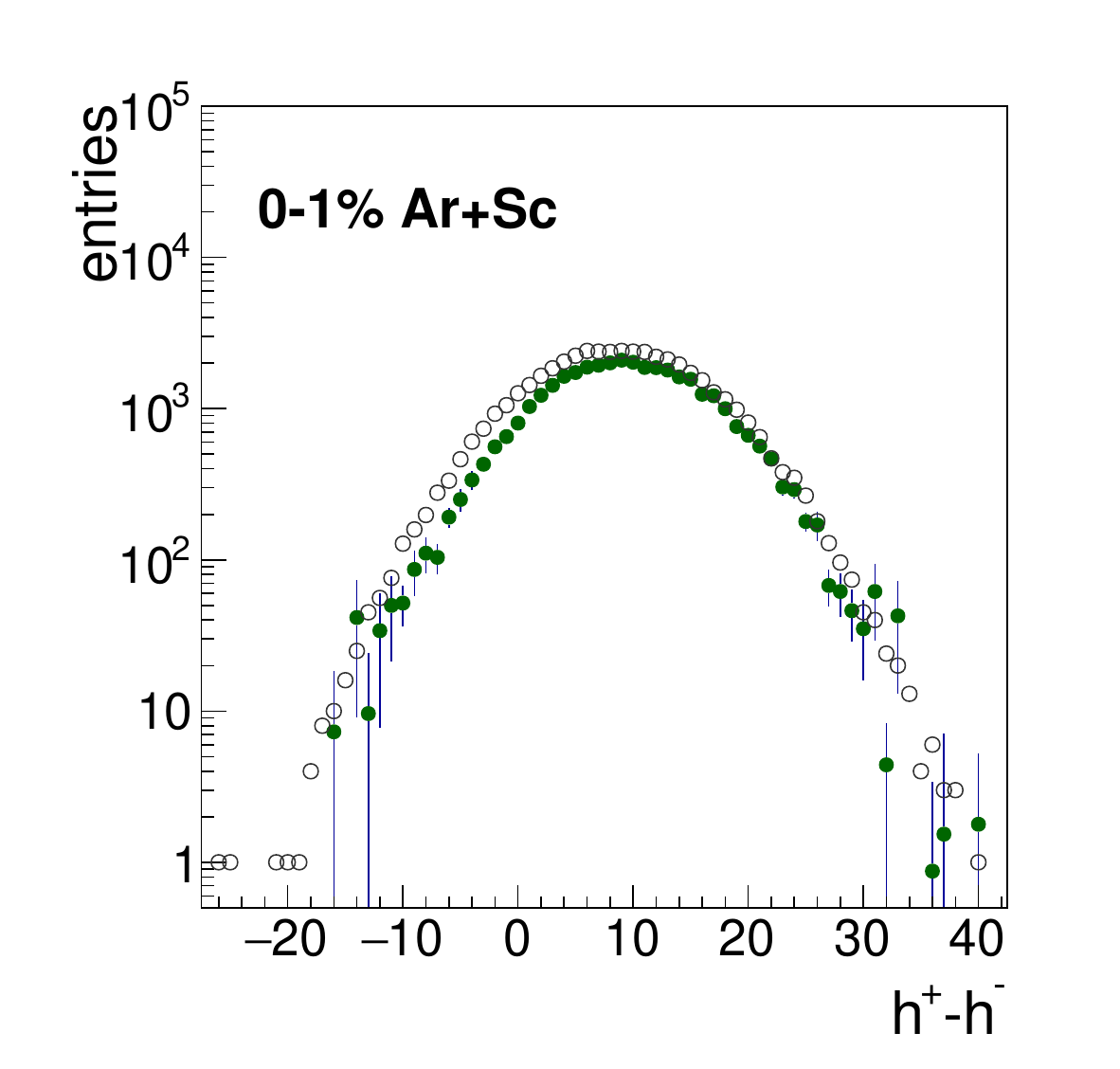}
    \end{minipage}
    \caption{The $E_{\rm{PSD}}$ distribution used for centrality selection in Ar+Sc collisions at 75\AGeVc (\textit{left}) and the net-charge distribution in the 1\% most central Ar+Sc collisions at 150\AGeVc (\textit{right}). The vertical line in the left panel indicates the selection threshold corresponding to the 1\% most central events (lowest $E_{\rm{PSD}}$). In the right panel, the open circles correspond to the measured distribution, whereas full circles indicate the corrected distribution.}
    \label{fig:tech}
    \end{figure}
Primary charged hadron distributions were measured in the 1\% most central Ar+Sc collisions, selected by identifying events with the lowest forward energy ($E_{\rm{PSD}}$), as measured by the Projectile Spectator Detector (PSD) calorimeter. An example of $E_{\rm{PSD}}$ for Ar+Sc interactions at 75\AGeVc is shown in the left panel of Fig.~\ref{fig:tech}.  Next, the hadron distributions were corrected for a possible bias of secondary or off-time interactions and weak decays using the unfolding method through the RooUnfold framework~\cite{unfolding}. The example of the corrected and uncorrected net-charge distribution is shown in the right panel of Fig.~\ref{fig:tech}.
\begin{figure}
    \centering
    \includegraphics[width=0.32\textwidth]{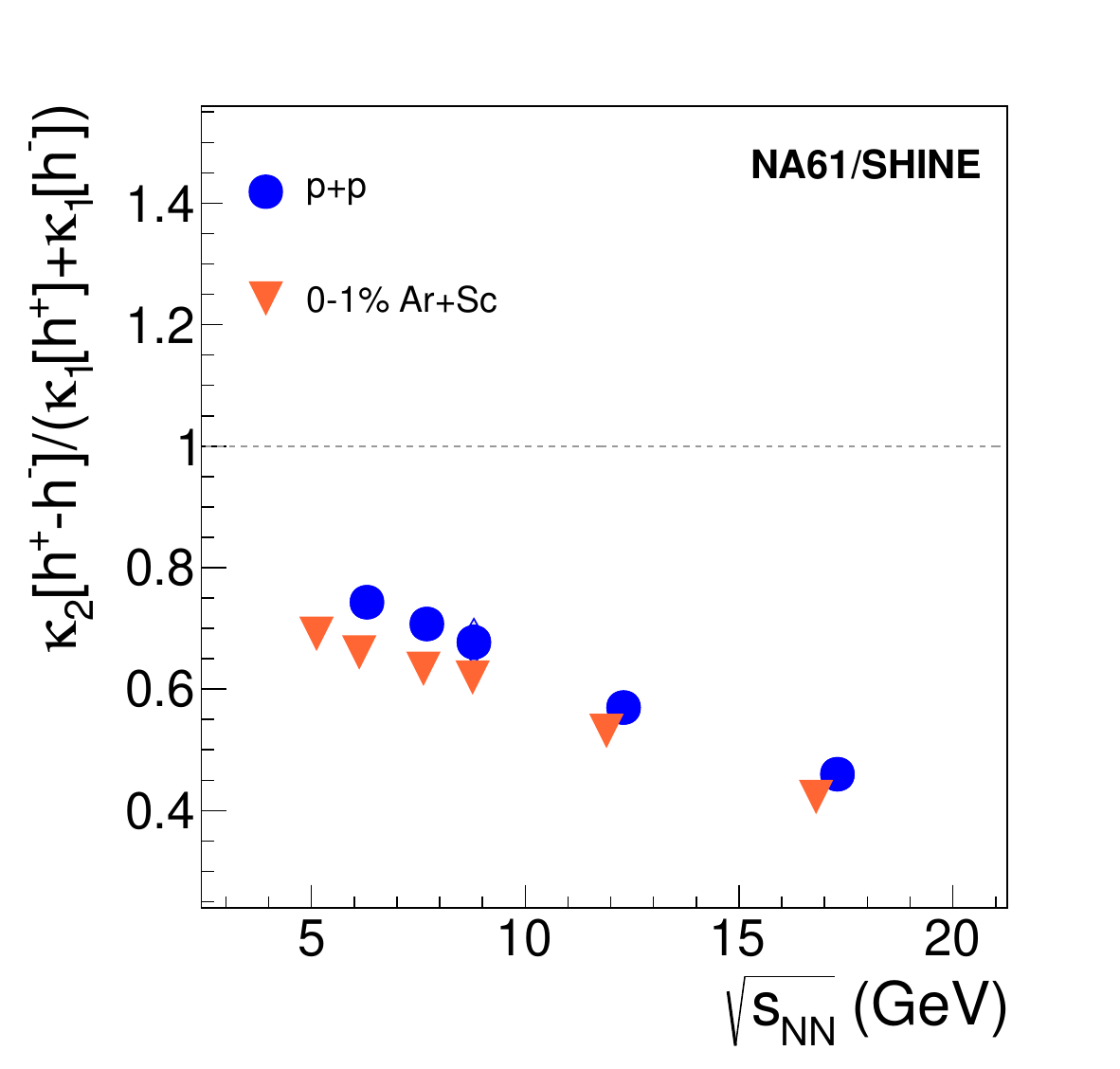}
    \includegraphics[width=0.32\textwidth]{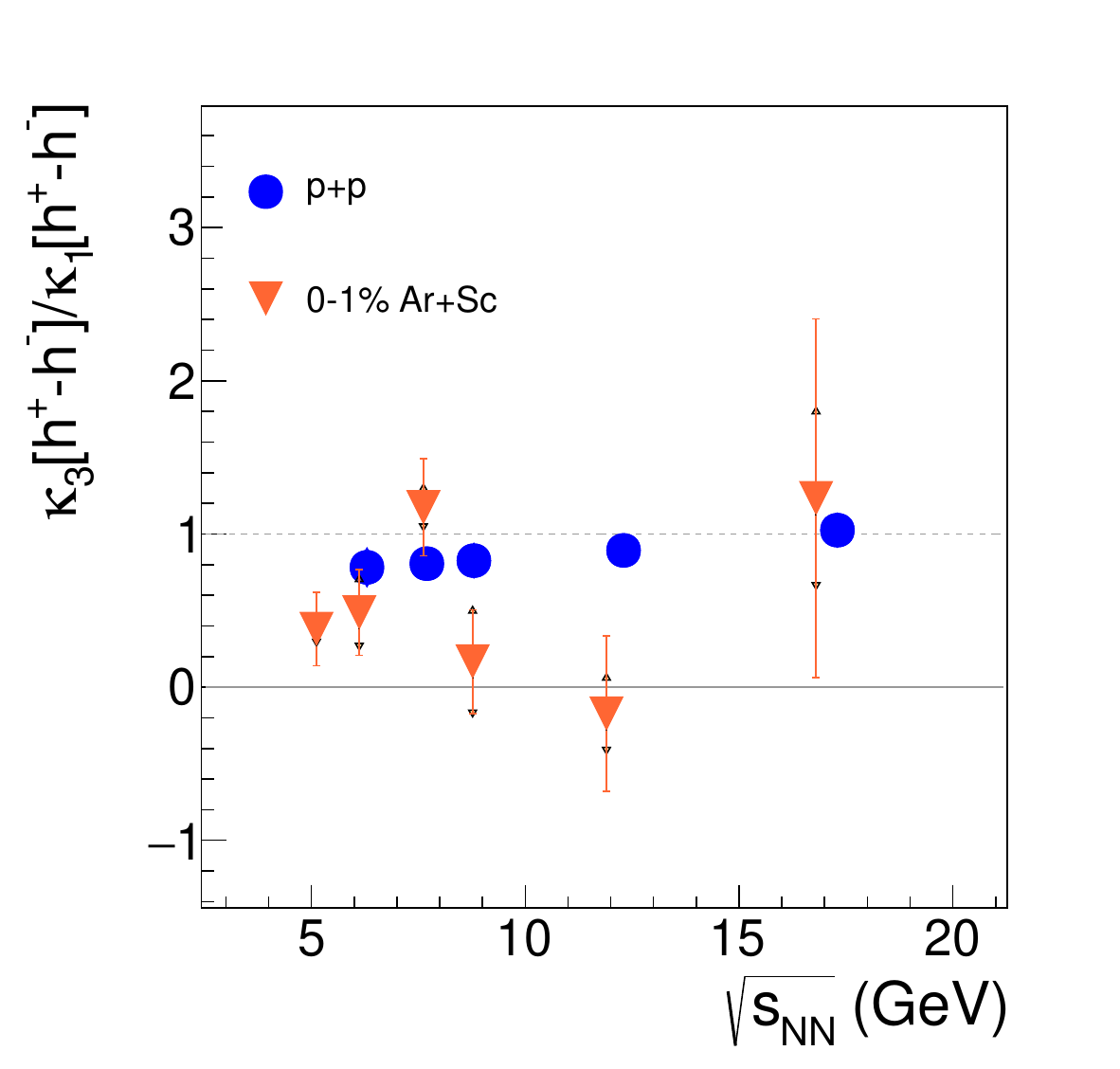}
    \includegraphics[width=0.32\textwidth]{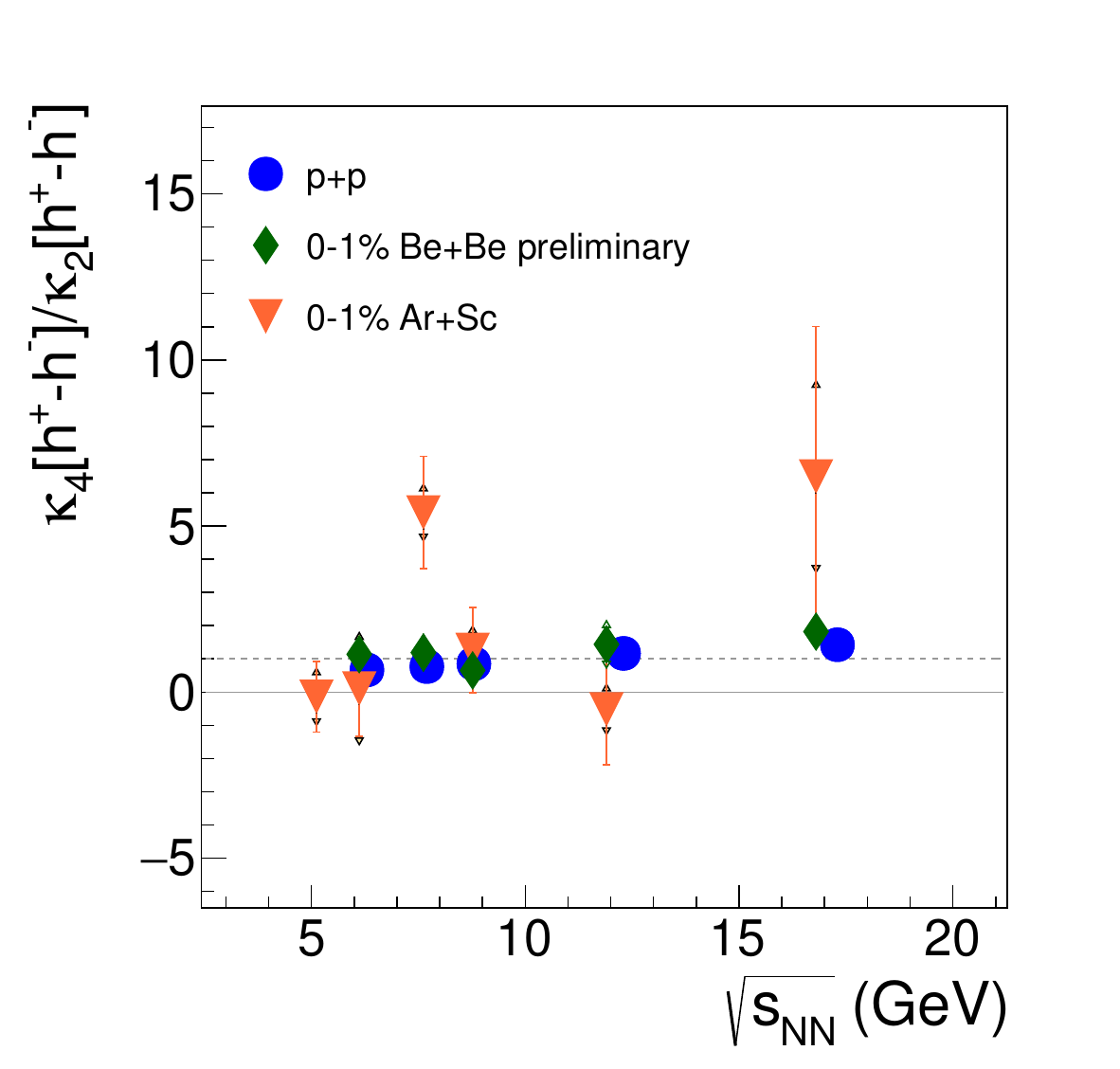}
    \caption{Energy dependence of intensive quantities of net-charge fluctuations in Ar+Sc collisions~\cite{NA61SHINE:2025whi} compared to published \pp~\cite{SHINE:2023ejm} and preliminary Be+Be results~\cite{Mackowiak-Pawlowska:2020glz}. 
Vertical bars denote statistical uncertainties, and vertical arrows -- systematic ones. Gray lines indicate the reference values.}
    \label{fig:results_net}
\end{figure} 
To eliminate volume dependence inherent in raw moments, ratios of cumulants are employed, as they provide greater sensitivity to critical phenomena~\cite{Stephanov:2008qz}.  
The measures are intensive and have two reference values: 0 for no fluctuations and 1 for the Skellam distribution (in case of multiplicity distribution, the Poisson distribution). In case of net-charge, $\kappa_{2}[h^{+}-h^{-}]/(\kappa_{1}[h^{+}]+\kappa_{1}[h^{-}])$, $\kappa_{3}/\kappa_{1}[h^{+}-h^{-}]$, and $\kappa_{4}/\kappa_{2}[h^{+}-h^{-}]$ were studied. 
For multiplicity fluctuation studies, the quantities are differently defined to keep the same reference values. The results on net-charge are shown in Fig.~\ref{fig:results_net} and compared to \NASixtyOne published \pp~\cite{SHINE:2023ejm} and preliminary Be+Be~\cite{Mackowiak-Pawlowska:2020glz} results within the same analysis acceptance~\cite{Detector_acceptance}. A hint of non-monotonic behavior is observed for the ratios $\kappa_{3}/\kappa_{1}$ and $\kappa_{4}/\kappa_{2}$ in Ar+Sc collisions. However, substantial statistical uncertainties make drawing any definitive conclusions difficult. In the case of multiplicity fluctuations, the largest difference between Ar+Sc and \pp interactions is observed for the second-order cumulant ratio ($\kappa_{2}/\kappa_{1}$) as shown in Fig.~\ref{fig:results_pos}. Smaller deviations of higher-order cumulant ratios can be observed at lower and mid-SPS energies -- for details, see Ref.~\cite{NA61SHINE:2025whi}. 

\begin{figure}
    \centering
    \includegraphics[width=0.32\textwidth]{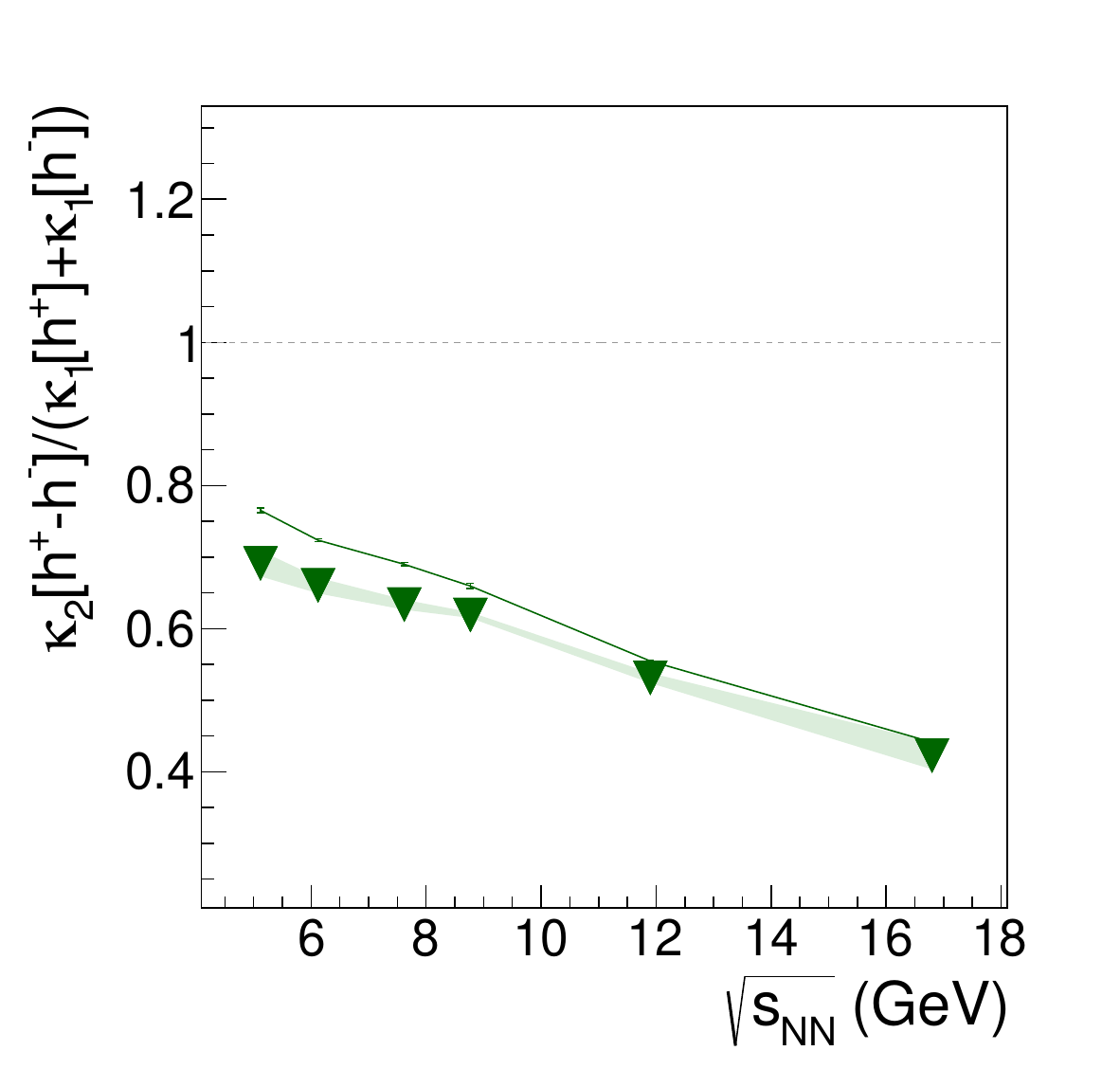}
    \includegraphics[width=0.32\textwidth]{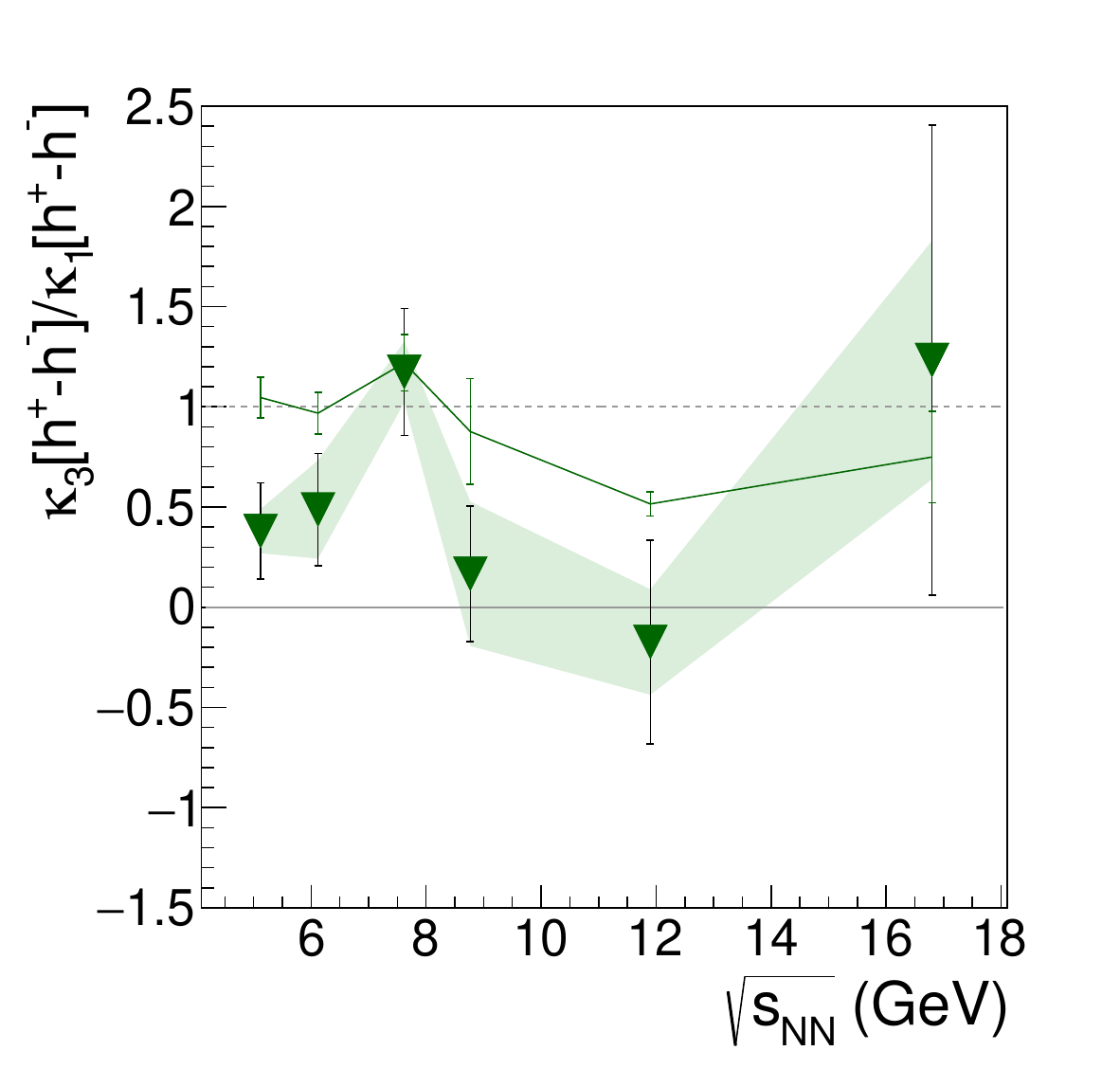}
    \includegraphics[width=0.32\textwidth]{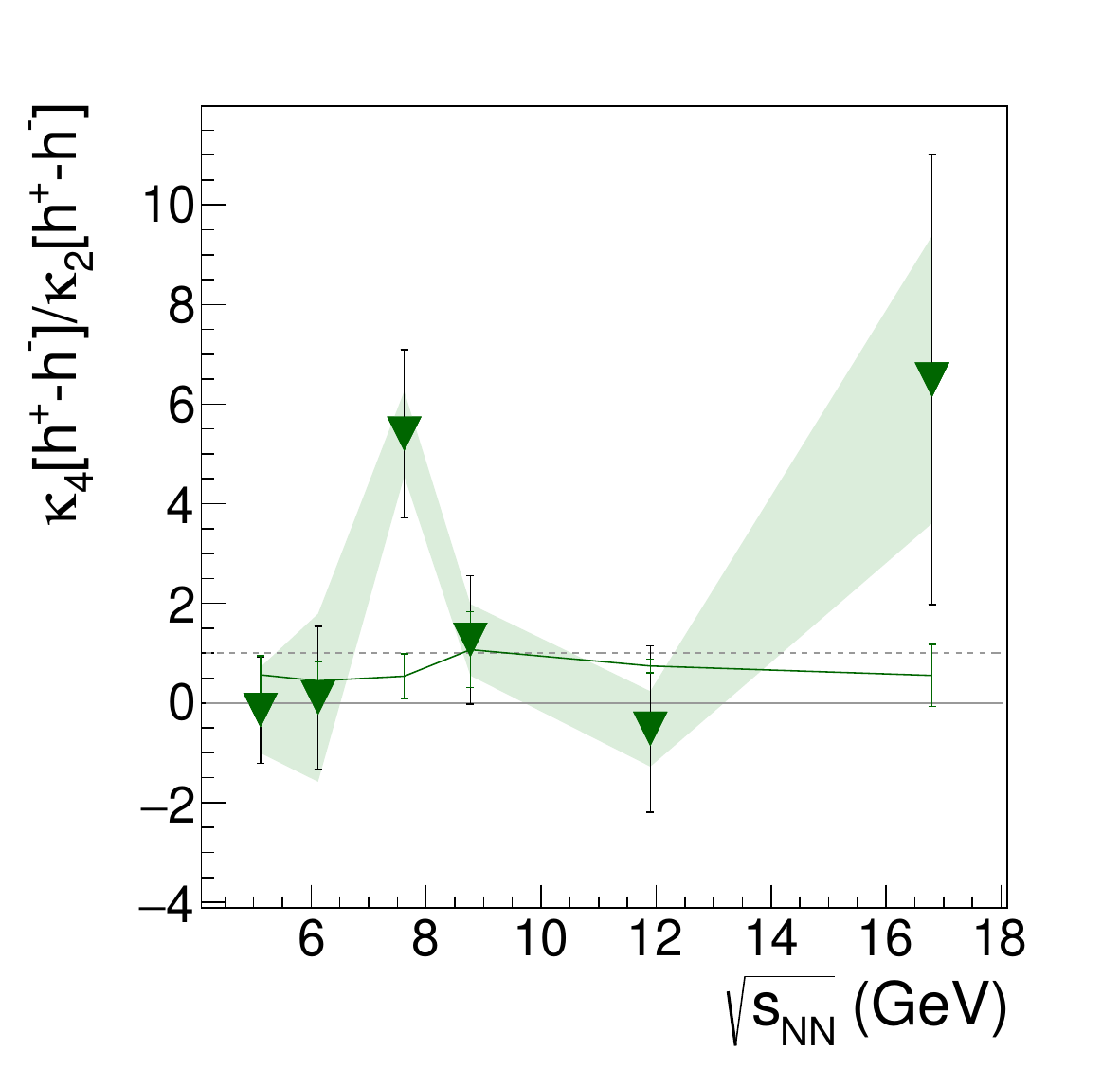}
    \caption{Comparison of the energy dependence of intensive quantities of net-charge distribution in Ar+Sc collisions (triangles) with EPOS1.99 model (line). Vertical bars denote statistical uncertainties, and color bands -- systematic ones. Gray lines indicate the reference values.}
    \label{fig:model}
\end{figure}

The comparison with the EPOS1.99 model~\cite{Werner:2008zza} is presented in Fig.~\ref{fig:model}. The model reproduces the magnitude of the signal in the data, although the most significant discrepancies appear at lower and mid-SPS energies. 
In the case of multiplicity fluctuations, the signal is generally smaller than 1 (see Ref.~\cite{NA61SHINE:2025whi}). The EPOS1.99 model reproduces the magnitude of the signal but does not qualitatively agree with it.
The observed non-monotonic structures in Ar+Sc collisions may be related to the onset of deconfinement~\cite{Gazdzicki:2003bb, Gorenstein:2003hk} or a critical point~\cite{Sarkar:2025xpv}, but additional evidence is necessary. Further studies in Ar+Sc collisions require either increasing data statistics or finding more suitable quantities, e.g. see Ref.~\cite{Sangaline:2015bma}.

\vspace{0.3cm}
\noindent
\small {\textbf{Acknowledgements:} This work was partially supported by the Polish Ministry of Science and Higher Education (grant WUT ID-UB)}.
\vspace{0.5cm}
\bibliography{references}

%
%

%
%

\end{document}